\documentclass[preprint,preprintnumbers,aps]{revtex4}

\usepackage[dvips]{graphicx}
\usepackage{graphicx}
\usepackage{amsfonts}
\usepackage{bm}
\usepackage{amsmath}
\usepackage{amssymb}
\usepackage{color}
\usepackage[all]{xy}
\usepackage{mathtools}
\usepackage{soul}
\usepackage{booktabs}

\usepackage{float}
\usepackage{subcaption}

\def\be{\begin{equation}}
\def\ee{\end{equation}}
\def\bea{\begin{eqnarray}}
\def\eea{\end{eqnarray}}

\usepackage[dvipsnames]{xcolor}

\begin{document}

\title{ Solving Einstein's Vacuum Equations with Physics-Informed Neural Networks: Boundary Conditions and Domain Decomposition}

\author{Elly Bayona$^1$ and Hernando Quevedo$^{1,2,3}$}
\email{elly.bayona@correo.nucleares.unam.mx,quevedo@nucleares.unam.mx}
\affiliation{$^1$Instituto de Ciencias Nucleares, Universidad Nacional Aut\'onoma de M\'exico, México.}
\affiliation{$^2$Dipartimento di Fisica and Icra, Universit\`a di Roma “La Sapienza”, Roma, Italy.}
\affiliation{$^3$ Al-Farabi Kazakh National University, Al-Farabi av. 71, 050040 Almaty, Kazakhstan.}

Corresponding  author: Elly Bayona (email: elly.bayona@correo.nucleares.unam.mx)

\newcommand{\figures}[5]{\begin{figure}[#1]
    \centering
    \includegraphics[width=#2\linewidth]{AxiSymmetricModel_Batches_ArtVrs3_#3_#4}
    \caption{#5}
    \label{fig:#3-#4}
\end{figure}}

\newcommand{\labels}[2]{fig:#1-#2}

\date{\today}


\begin{abstract}

We investigate the application of Physics-Informed Neural Networks (PINNs) to the numerical solution of Einstein's vacuum field equations for static spacetimes. We first reproduce the Schwarzschild solution and then extend the method to the axisymmetric $q-$metric, a nontrivial exact solution characterized by a mass quadrupole moment. We analyze the influence of boundary conditions, domain decomposition, and equation redundancy on the convergence and stability of the training process. The proposed framework accurately reconstructs the metric functions in the computational domain while maintaining small residual errors. Our results demonstrate that PINNs provide a robust and flexible approach to solving Einstein's equations and offer a promising foundation for investigating gravitational configurations for which exact analytical solutions are unknown.


\end{abstract}


\maketitle


\section{Introduction}
\label{sec:int}

Machine Learning (ML) has become an important tool in modern scientific research, giving rise to the rapidly growing field of Scientific Machine Learning (SciML), whose main objective is to integrate data-driven techniques with the governing laws of physical systems \cite{Karniadakis2021SciML}, \cite{Cuomo2022ReviewPINN}. Among the different approaches that have emerged in this context, Physics-Informed Neural Networks (PINNs) \cite{Raissi2019PINN} have attracted particular attention because they incorporate the differential equations of the underlying theory directly into the training process. As a consequence, they provide continuous approximations of the solutions without requiring traditional mesh-based discretization methods.

PINNs have been successfully applied to a broad range of problems involving ordinary and partial differential equations in fluid dynamics, plasma physics, electromagnetism, astrophysics, and cosmology. In general relativity, they offer an attractive alternative for solving Einstein's field equations, whose nonlinear structure often makes analytical solutions impossible and numerical calculations computationally demanding \cite{li2023solvingSchw,stefanou2023pulsar,QNSchw}. Rather than replacing traditional numerical relativity methods, PINNs provide a complementary framework in which the field equations themselves become the central ingredient of the optimization process.

Despite the increasing interest in this subject, most existing applications of PINNs in general relativity have concentrated on reproducing known exact solutions or on assisting standard numerical algorithms. Consequently, several fundamental questions remain open. Can a neural network reliably solve Einstein's equations for genuinely axisymmetric spacetimes? What boundary conditions are required to guarantee convergence? How does the coordinate freedom of general relativity influence the optimization process? Does the redundancy of Einstein's equations improve or deteriorate the numerical performance? Answering these questions is essential before PINNs can be used to investigate gravitational configurations for which no analytical solutions are available.

In this work we address these questions by considering two vacuum spacetimes of increasing complexity. We first reproduce the Schwarzschild solution, which provides a well-understood benchmark for validating the numerical implementation and analyzing the influence of equation redundancy. We then consider the q-metric, the simplest exact static vacuum solution possessing a genuine quadrupole moment. Unlike Schwarzschild spacetime, the q-metric depends on two coordinates and involves four coupled metric functions, making it a significantly more demanding benchmark for PINNs. Moreover, because the q-metric continuously reduces to the Schwarzschild solution in the limit q=0, it provides a natural framework for investigating how the complexity of Einstein's equations affects the training process.

Our main objective is not simply to reproduce known exact metrics, but to investigate the numerical behavior of PINNs when solving Einstein's equations under different strategies for imposing boundary conditions and domain decomposition. We compare strict and relaxed boundary conditions, analyze the role of redundant Einstein equations, and study the stability and convergence of the training process. These issues are of direct relevance for future applications in which the exact spacetime geometry is unknown and PINNs may become a practical tool for constructing new numerical solutions of Einstein's equations.

The paper is organized as follows. In Sec. \ref{sec:solutions}, we review the main properties of the exact vacuum solutions we investigate in this work. 
Section \ref{sec:PINNs}
reviews the basic formulation of PINNs. Section \ref{sec:Schw}
III presents the Schwarzschild case and discusses the role of redundant equations. Section \ref{sec:qMetric}  analyzes the q-metric and the corresponding numerical strategy. Sections  \ref{sec:StrictConditions}  and \ref{sec:RelaxedConditions} analyze strict and relaxed boundary conditions, respectively. In Sec. \ref{sec:VM}, we validate the results obtained in the previous sections. Finally, Section \ref{sec:con} summarizes our conclusions and discusses future applications.

\section{Spherically and axially symmetric solutions}
\label{sec:solutions}

The only static, spherically symmetric, vacuum spacetime is described by the Schwarzschild line element, which can be  expressed as:
\begin{equation}\label{schwarszchild}
ds^2_{sch} = -\left(1-\frac{2m}{r}\right) dt^2 +\left(1-\frac{2m}{r}\right)^{-1}dr^2 +r^2d\Omega^2 \; ,
\end{equation}
where $d\Omega^2=d\theta^2+\sin^2\theta d{\varphi}^2$ is the line element of a 2-sphere. This metric describes the gravitational field of a black hole of mass $m$. It possesses a curvature singularity at $r=0$ and a coordinate singularity at $r=2m$, a hypersurface which is known as the event horizon.

The simplest spacetime that describes the exterior gravitational field of a static and axisymmetric mass distribution with only a quadrupole moment is determined by the quadrupolar metric (q-metric), which in Schwarzschild-like coordinates can be written as  \cite{qMetric_23} given by
\begin{align}
    ds^2=&-\left(1-\frac{2m}{r}\right)^{1+q}dt^2+\left(1-\frac{2m}{r}\right)^{-q}\nonumber\\
    &\times \left[\left(1+\frac{m^2\sin^2\theta}{r^2-2mr}\right)^{-q(q+2)}\left(r^2d\theta^2+\frac{dr^2}{1-2m/r}\right)+r^2\sin^2\theta\ d\varphi^2 \right]\ .\label{q}
\end{align}

The q-metric is obtained from the Schwarzschild metric by applying  a Zipoy-Voorhees transformation \cite{qMetric_24,qMetric_25}. The geometric and physical properties of this metric have been investigated in detail in  \cite{qMetric_21,qMetric_23,qMetric_25,qMetric_26,Quevedo2017Quadrupolar,QuevedoPedro,qMetric_28}. The  deformation of the gravitational source is described by the quadrupolar parameter $q$. In the limiting case $q=0$, the q-metric reduces to the Schwarzschild metric, whereas the limit $q=1
$ corresponds to the flat Minkowski spacetime. 
Moreover, the parameter $m$ is related to the total mass  of the gravity source. These properties follow from the analysis of the corresponding Geroch \cite{geroch1970multipole1,geroch1970multipole2} relativistic  multipole moments, $M_n \ (n=0,1,...) $, which  are given by 
\begin{align}
    &M_0=m(q+1)\ ,& &M_1=0\ , & &M_2=-\frac{m^3}{3}q(q+1)(q+2),\label{momentosMul}\\
    &M_{2k+1}=0\ ,& &M_{2k}\sim q+1\ , & &  k=0,1,\ldots
\end{align}

An essential feature of this spacetime is that   the hypersurface $r=2m$ is no longer a coordinate singularity, but a physical spacetime singularity. Consequently, in general, it describes the exterior gravitational field of a naked singularity  \cite{qMetric_23}. Furthermore, the q-metric can be matched with a class of perfect-fluid interior solutions with predefined density profiles  \cite{Quevedo2017Quadrupolar} such that the naked singularity becomes covered by the interior configuration. 

The q-metric constitutes an ideal model for testing PINNs because it represents the simplest exact axisymmetric vacuum solution beyond spherical symmetry. In contrast to the Schwarzschild spacetime, it depends on two coordinates and four coupled metric functions while remaining analytically known, allowing a direct quantitative comparison between the numerical and exact solutions.


\section{Physics-informed neural networks (PINNs)}
\label{sec:PINNs}

Physics-informed neural networks (PINNs) are a SciML technique that incorporates physical laws, usually formulated as partial differential equations (PDEs), directly into the learning process of a neural network. The fundamental difference from typical network models is that they do not require data as a fundamental part of the training process, since the PDEs replace this function. However, observational data can be incorporated into networks in a technique known as Inverse Physics-Informed Neural Networks (I-PINNs) and usually used when the goal is to infer physical parameters of the system \cite{Karniadakis2021SciML}.

In inversion problems, it is possible to incorporate these parameters into the training process to estimate them. However, the problems considered in this work are of the forward type, so these parameters remain fixed.

Let a second-order PDE be defined on a domain $\Omega\subset \mathbb{R}^n$ and depend on a set of fixed parameters $\boldsymbol{\lambda}$.  The equation governing the system is given by
\begin{equation}
    \mathcal{F}\left(\hat{x},u(\hat{x}),\nabla u(\hat{x}), \nabla^2 u(\hat{x});\boldsymbol{\lambda} \right)=0\ ,
\end{equation}
where the coordinates are denoted by $\hat{x}=(x_1,\ldots,x_n)\in \Omega$ and $u(\hat{x})$ is the exact solution of the system evaluated at $\hat{x}$. Associated with this equation, the boundary conditions on $\partial\Omega$ are given by
\begin{equation}
    \mathcal{B}(u,\hat{x})=0\ ,\qquad \hat{x}\in\partial\Omega\ .
\end{equation}

\medskip

The training process produces an output function $u_{\text{NN}}(\hat{x};\theta)$ that approximates the exact solution of the differential equation, namely,
\begin{equation}
u_{\text{NN}}(\hat{x};\theta)\approx u(\hat{x})\ .
\end{equation}
Here, $\theta$ denotes the internal trainable parameters of the NN. These parameters are determined by minimizing a loss function $\mathcal{L}(\theta)$, according to
\begin{equation}
    \theta= \arg\min_{\theta} \mathcal{L}(\theta)\ .
\end{equation}
In the context of PINNs, the loss function is generally constructed as the sum of three main contributions. The first corresponds to the PDE residual loss and is defined as
\begin{equation}
    \mathcal{L}_\mathcal{F}(\theta) = \|\mathcal{F}\left(\hat{x},
    u_{\text{NN}}(\hat{x};\theta),
    \nabla u_{\text{NN}}(\hat{x};\theta),
    \nabla^2 u_{\text{NN}}(\hat{x};\theta);\boldsymbol{\lambda}
    \right)\|^2
\end{equation}  
where $\|\cdot\|$ denotes a discrete norm evaluated over the set of $N_f$ collocation points at which the PDE residual is computed. In this work, the $L_2$ norm is employed. Therefore, the differential equation is evaluated using the neural-network approximation $u_{\text{NN}}$ and the corresponding PDE residual is minimized over the set of the collocation points sampled within the domain.
The second contribution of the loss function enforces the boundary conditions on a set of $N_b$ points over the boundary $x_b \in \partial\Omega$ and is given by, 
\begin{equation}
    \mathcal{L}_\mathcal{B}(\theta) = \|\mathcal{B}\left(\hat{x}_b,u_{\text{NN}}(\hat{x}_b;\theta)\right)\|^2\ .
\end{equation}      
Finally, some problems, such as inverse problems, incorporate experimental data or observational data $(\hat{x}_j,u_j)$. Such information can be included in the training process through the following loss term
\begin{equation}
\mathcal{L}_{\text{data}}(\theta)
= \bigl\| u_{\text{NN}}(\hat{x}_j;\theta) - u_j \bigr\|^2\ .
\end{equation}
The total loss function is therefore given by,
\begin{equation}\label{pesosdePerdida}
    \mathcal{L}(\theta) =\omega_\mathcal{F}\mathcal{L}_\mathcal{F}(\theta) + \omega_\mathcal{B}\mathcal{L}_\mathcal{B}(\theta) + \omega_d\mathcal{L}_{data}(\theta) \ ,
\end{equation}
where $\omega_\mathcal{F}$, $\omega_\mathcal{B}$, and $\omega_{d}$ are hyperparameters used to balance the different contributions to the loss function and to improve the stability of the training process. Since the present work focuses on vacuum solutions of Einstein's equations without observational or synthetic data, the data term is omitted throughout the numerical implementation, i.e., we set $\omega_d=0$. This term  is included here only for completeness because it constitutes the standard PINN formulation.

\begin{figure}[H]
    \centering
    \includegraphics[width=1.\linewidth]{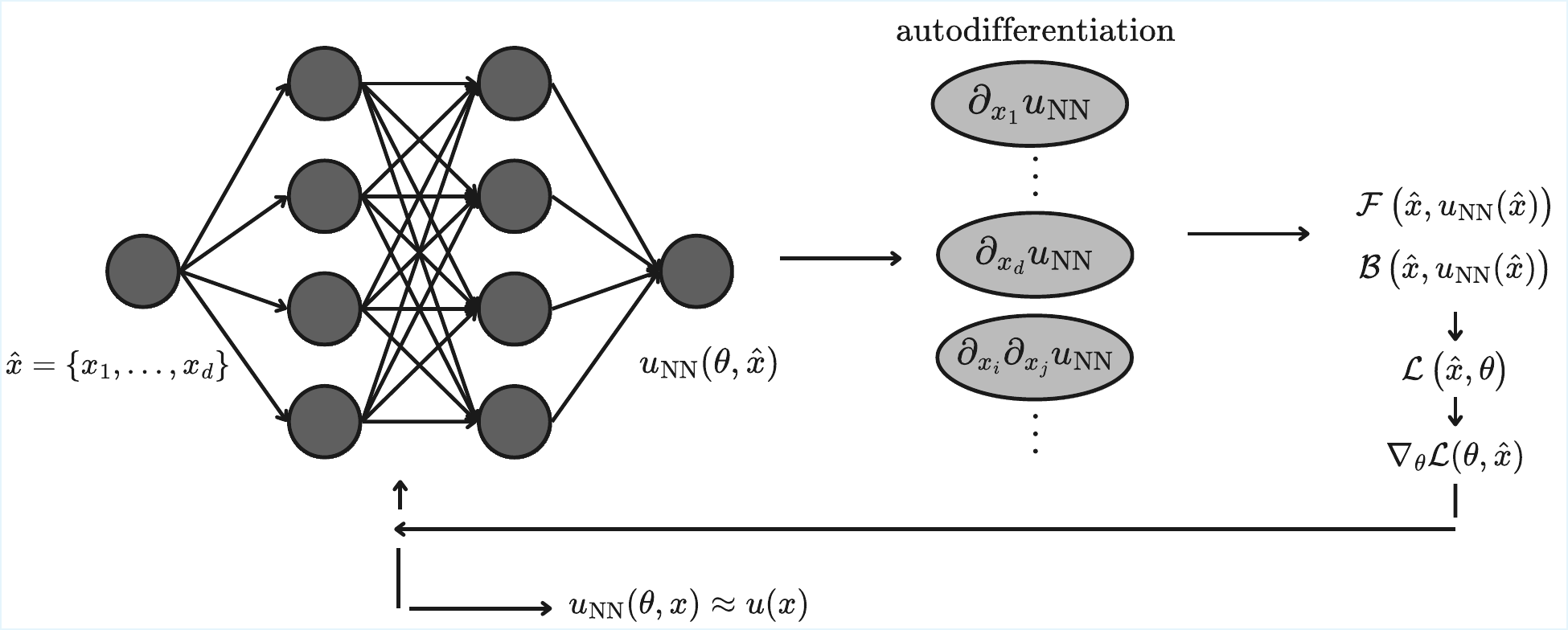}
    \caption{Flowchart of the training of a Physics-Informed Neural Network PINNs, where partial differential equations are directly incorporated into the training process. Adapted from \cite{li2023solvingSchw}.}
    \label{PINNs}
\end{figure}

The gradient of the loss function is computed with respect to all trainable parameters of the network, including its weights and biases. Figure \ref{PINNs} illustrates the training process used in this class of networks. The required partial derivatives are computed using automatic differentiation technique, which allows for the accurate evaluation of derivatives directly from the computational graph.
Automatic differentiation is particularly advantageous in general relativity because Einstein's equations require first- and second-order derivatives of the metric functions, which can be evaluated accurately without finite-difference approximations \cite{Griewank2008,Bradbury2018JAX}.


\section{Numerical PINN solution for spherical symmetry}
\label{sec:Schw}

To illustrate how the PINNs method works in the context of general relativity theory, we use it to solve the differential equations coming from the Einstein field equations over a spherically symmetric vacuum spacetime, namely, the case of the Schwarzschild spacetime presented in Eq.~\eqref{schwarszchild}. 
The Schwarzschild spacetime provides the simplest nontrivial vacuum solution of Einstein's equations and therefore constitutes an ideal example for validating the PINN implementation before addressing genuinely axisymmetric geometries.

The analysis of the Schwarzschild metric with PINNs was previously developed in \cite{li2023solvingSchw}, where several NNs were constructed to solve Einstein's field equations. In particular, for the vacuum case, the tensor equation corresponds to $R_{\mu\nu}=0$. 
The approach adopted in \cite{li2023solvingSchw} consists in explicitly incorporating the field equations by computing the geometric quantities involved in the definition of the Ricci tensor, such as the Christoffel symbols and their derivatives. However, for practical reasons, in the present work these expressions are not analyzed and, instead, we focus directly on the field equations. The objective is to obtain results with error magnitudes comparable to those reported in the \cite{li2023solvingSchw}. Specifically, the neural network must satisfy the following three equations:

\begin{align}
    Eq_1\equiv&\frac{h_r}{2\sqrt{fg}}+\frac{f_r}{rfg}=0\\
    Eq_2\equiv&-\frac{h_r}{2\sqrt{fg}}+\frac{g_r}{rg^2}=0\\
    Eq_3\equiv&-\frac{f_r}{2rfg}+\frac{g_r}{2rg^2}+\frac{1}{r^2}-\frac{1}{r^2g}=0 \ ,  
\end{align}
where $f(r)$ and $g(r)$ are the two metric functions that define the general spherically symmetric line element given by,
\begin{equation}
    ds^2=-f(r) dt^2 +g(r) dr^2 +r^2 d\Omega^2\ .
\end{equation}
The two metric functions involved are related to the PINNs by means of two neural networks, $u_0(r)$ and $u_1(r)$, through the following expressions,
\begin{align}
    f(r)&=1-\frac{2m}{r}-\frac{u_0(r)}{r^2}\ ,\label{eq.fSchw}\\
    g(r)&=u_1(r)\; .
\end{align}
As mentioned in \cite{li2023solvingSchw}, the function $f(r)$ incorporates the expected asymptotic behavior, satisfying the boundary condition where $f(r)\rightarrow 1-2m/r$ when $r\rightarrow \infty$. The loss function (\emph{loss}) is defined from the residuals of the previous differential equations evaluated over a set of training points and by averaging the squares of the aforementioned residuals. In particular, if the subscript $i=\{1,2,3\}$ denotes the three differential equations considered and the subscript $j$ runs over the $N$ points of the collocation set in the domain, the expression used is
\begin{equation}
    \mathcal{L}=\frac{1}{3N}\sum_{i=1}^{3}\sum_{j=1}^N \omega_i(Eq_i(r_j))^2
\end{equation}
where $\omega_i$ are weights associated with the residual of each differential equation.

\medskip

\begin{figure}[H]
    \centering
    \begin{subfigure}[b]{0.48\linewidth}
        \centering
        \includegraphics[width=\linewidth]{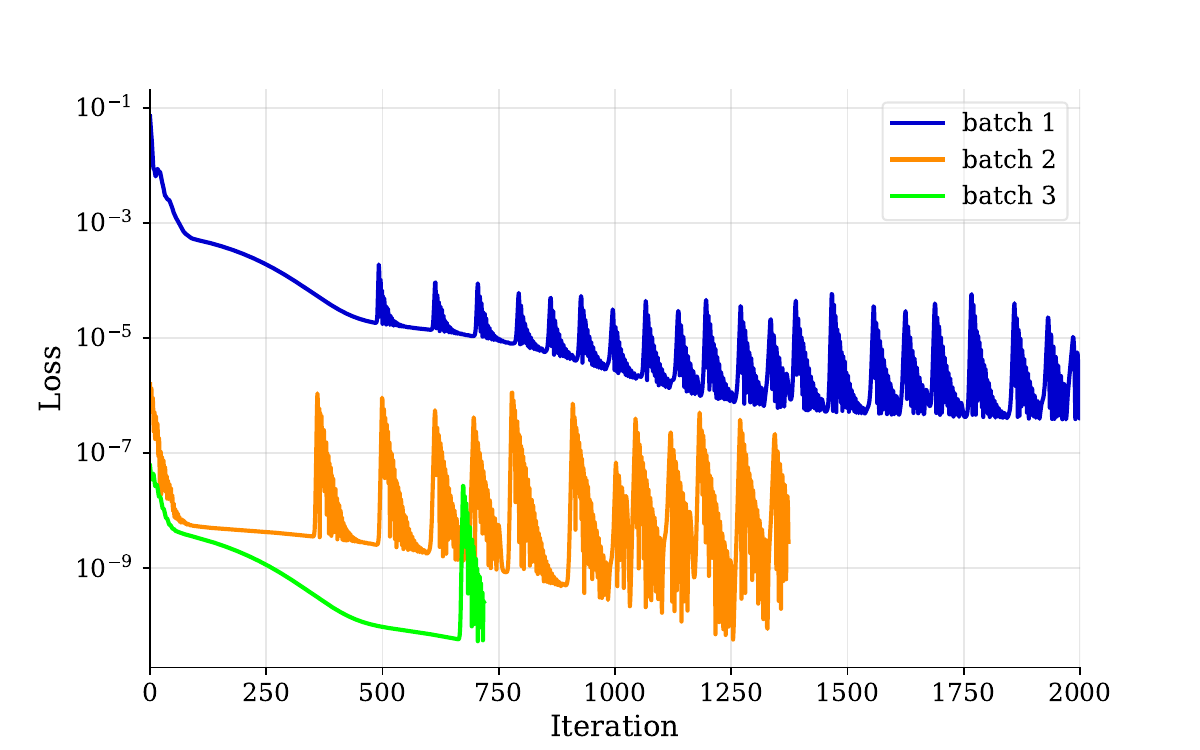}
        \caption{Total loss function.}
        \label{fig:Schw_total_batches}
    \end{subfigure}
    \hfill
    \begin{subfigure}[b]{0.48\linewidth}
        \centering
        \includegraphics[width=\linewidth]{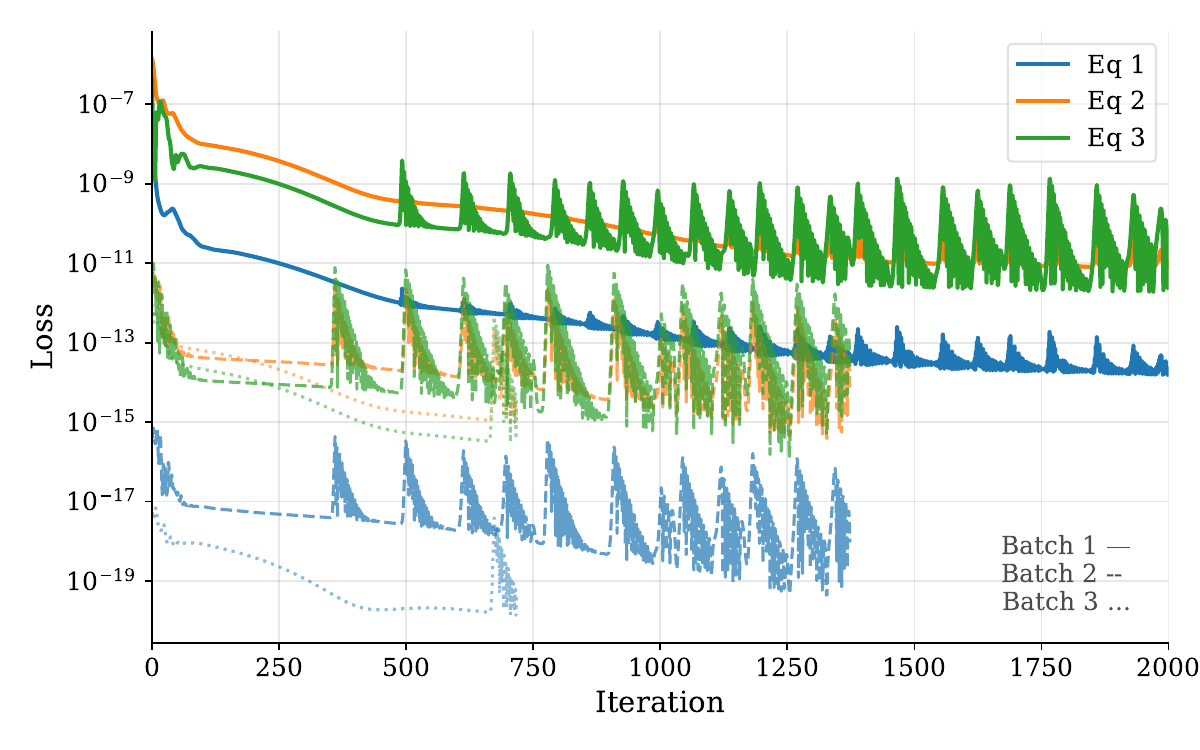}
        \caption{Loss function per equation.}
        \label{fig:Schw_eqs_batches}
    \end{subfigure}
    
    \caption{Convergence of the loss functions for Schwarzschild spacetime using a domain decomposition technique.}
    \label{fig:Schw_combined_batches}
\end{figure}
\begin{figure}[H]
    \centering
    \begin{subfigure}[b]{0.48\linewidth}
        \centering
        \includegraphics[width=\linewidth]{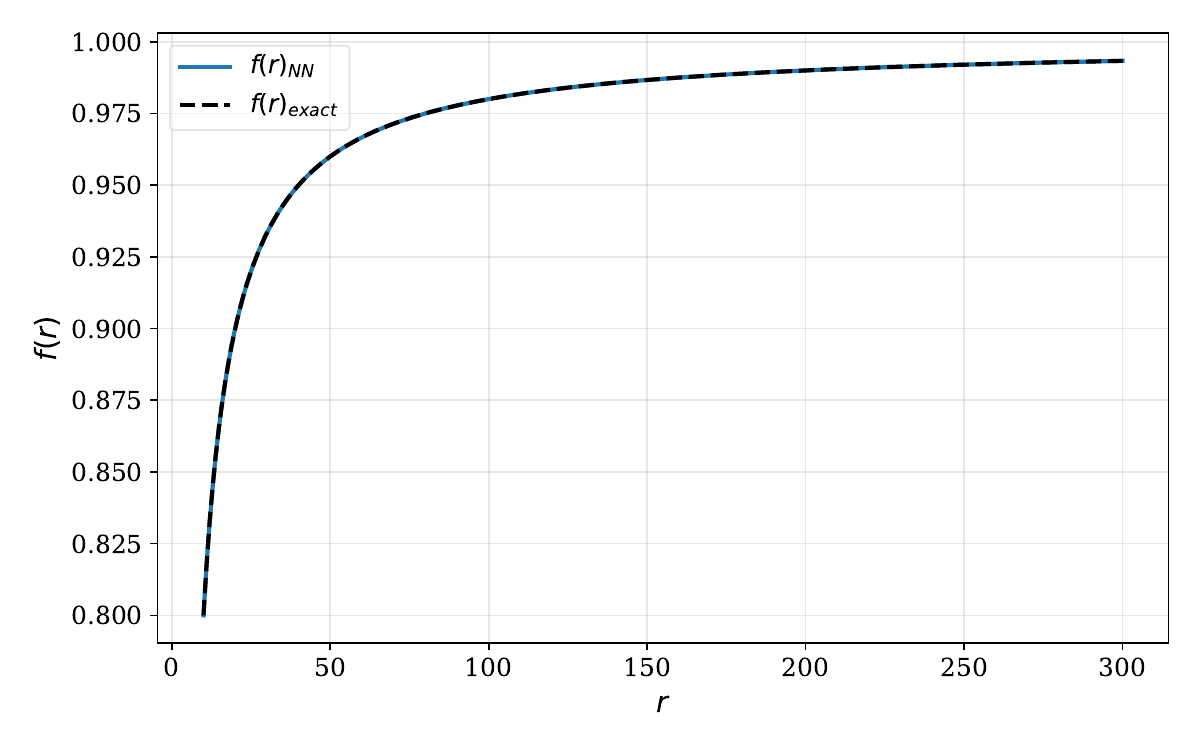}
        \caption{Metric function $f(r)$.}
        \label{fig:eqF_batches}
    \end{subfigure}
    \hfill
    \begin{subfigure}[b]{0.48\linewidth}
        \centering
        \includegraphics[width=\linewidth]{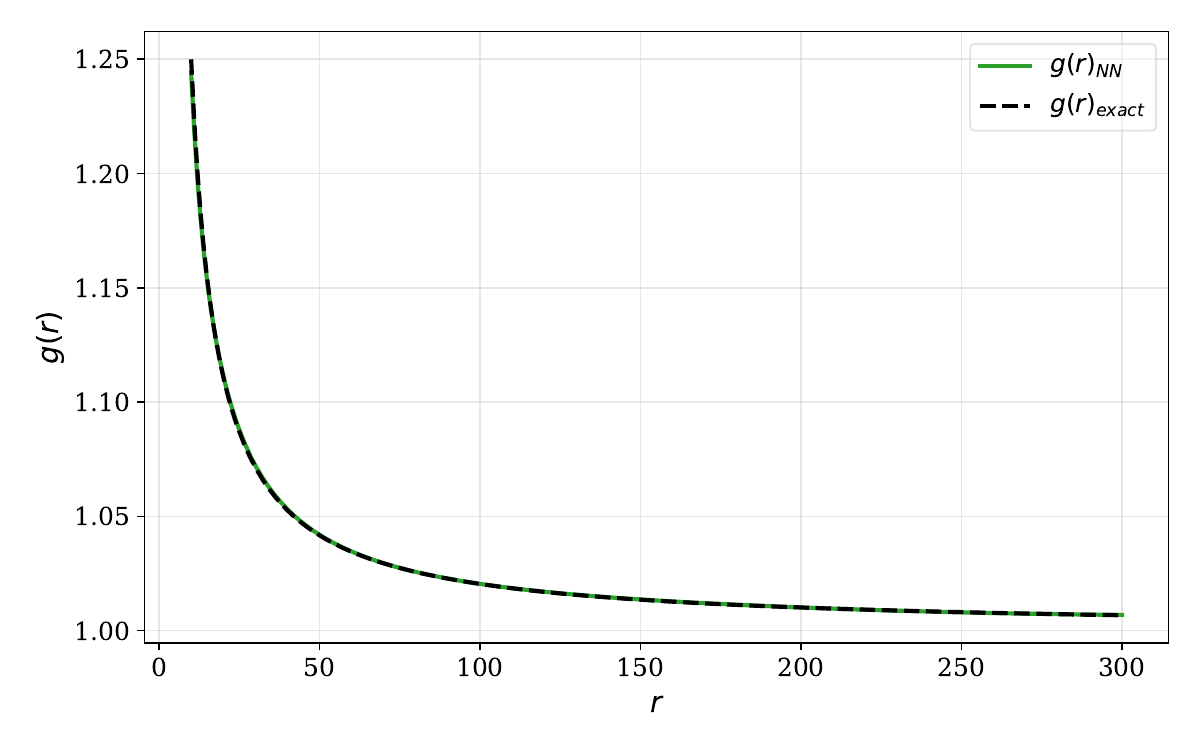}
        \caption{Metric function $g(r)$.}
        \label{fig:eqG_batches}
    \end{subfigure}
    
    \caption{Comparison between the exact solution and the PINN solution obtained using domain decomposition.}
    \label{fig:Comparacion_batches}
\end{figure}

The results are shown in Fig.~\ref{fig:Schw_combined_batches} and Fig.~\ref{fig:Comparacion_batches}. They were obtained after 2000 training iterations using two independent neural networks, each composed of 64 neurons. We adopted a domain decomposition technique. Then, the radial domain, $r=(10m,300m)$, was divided into three intervals, $(10m,30m)$, $(30m,80m)$, and $(80m,300m)$ and trained sequentially. The weights obtained in each interval were used to initialize the training of the next one, allowing the information to propagate progressively from the inner to the outer regions. This procedure improves training stability across the entire domain.

\medskip

The {\tt Adam} optimizer is used with exponential decay coefficients {\tt betas=(0.9,0.99)}. The learning rate is initially set to {\tt Learning Rate=$10^{-3}$} and reduced by $10\%$ every 1000 iterations using an exponential decay scheme. 
The model employs {\tt LogSigmoid} activation function in the hidden layers and {\tt Softplus} in the output layer, allowing the introduction of nonlinearity in the hidden layers while enforcing positivity of the output function. 
Following the referenced article, the weights are initialized from a uniform distribution in the interval $(-1/\sqrt{n_0},1/\sqrt{n_0})$, where $n_0$ denotes the number of neurons in the output layer. 

\medskip

The two independent neural networks used to obtain the metric functions have separate optimizers and contribute to the overall loss function. This approach is adopted for computational simplicity, while a fully coupled formulation is reserved for the case of the q metric, considered in the following sections, which is computationally more demanding. 

\medskip


The total loss function is showed in Fig.~\ref{fig:Schw_total_batches} for each batch, while the loss per equation and per batch is shown in Fig.~\ref{fig:Schw_eqs_batches}. Although the loss function oscillates during the training, the procedure is numerically stable because the loss remains bounded and exhibits a decreasing overall behavior, eventually reaching the regime dominated by numerical errors. It is noted that the first batch exhibits greater difficulty during training. Consequently, approximately $1\times10^4$ iterations are required to reach a loss on the order of $3.5\times10^{-8}$. However, thanks to the sequential training scheme, the second batch requires approximately 1500 iterations to reach a loss on the order of $5\times10^{-11}$, while the third reaches this same level after approximately 1000 iterations. This is because training automatically stops when the loss function falls below a fixed threshold of $5\times10^{-11}$.

Figure~\ref{fig:Comparacion_batches} presents the comparison between the exact solutions and the approximate solutions generated by the neural network for each metric function. The use of batch training leads to a significant improvement in accuracy across the extended radial domain, compared to training performed on a single domain. 

\medskip

Finally, to establish the accuracy of the obtained solution, we calculate the Relative Root Mean Squared Error (RMSE) for the metric functions $f(r)$ and $g(r)$. The results are presented in Table~\ref{tab:RMSE_schw}. The PINN approximation for the metric function $f(r)$ achieves a relative RMSE on the order of $10^{-8}$, while the approximation for $g(r)$ yields a relative RMSE of approximately $10^{-4}$. Although the accuracy of $g(r)$ is lower, it still provides a good approximation to the exact solution corresponding to an average relative error of $0.01\%$.  Therefore, these results support the use of PINNs as an interesting approach for solving Einstein's field equations in more general settings.

\begin{table}[H]
    \centering
    \begin{tabular}{ccc}
         Metric function &  $f(r)$ & $g(r)$  \\
                \hline
         RMSE   & $\qquad 8.63\times10^{-9}\qquad $ & $7.10\times 10^{-4} $ 
    \end{tabular}
    \caption{Relative Root Mean Squared Error (RMSE) for the two metric functions for the Schwarzschild spacetime.}
    \label{tab:RMSE_schw}
\end{table}

\subsection{Independent Set of equations}

Although the Schwarzschild spacetime is described by two metric functions, the field equations generate three non-trivial equations, $R_{tt}=0$, $R_{rr}=0$ and $R_{\theta\theta}=0$. These equations are not independent since the Bianchi identities ensure that if $R_{tt}=0$ and $R_{rr}=0$ are satisfied, the last equation $R_{\theta\theta}=0$ is automatically satisfied. In principle, the network should be able to handle any number of differential equations, as long as the system is consistent, regardless of whether the equations form an independent set. 

\medskip

Nevertheless, it is worth investigating whether using an independent set of differential equations can improve the network performance. Not using such a set may influence the training process by introducing numerical conflicts during optimization. 
In traditional numerical methods, it is well known that small violations of redundant equations can propagate and increase the errors in the remaining equations \cite{alcubierre2008introduction}. In the context of PINNs, this could mean that the residuals of the additional equations could obstruct or block the training progress.
However, because PINNs methods are a relatively recent area of research, it is not yet clear whether these phenomena manifest themselves in a strictly analogous way to classical numerical methods. For example, in  \cite{li2023solvingSchw}, the model is trained using the full set of Ricci tensor equations, without examining the independence of the differential equations, and nevertheless obtains excellent results.

Given that the problems addressed in this work become progressively more complex, the behavior of the network is evaluated using the complete set of differential equations, whenever the system allows it. Furthermore, it is important to note that this effect can, in principle, be controlled through an appropriate choice of weights in the loss function.



\section{Numerical PINN solution for q-metric}
\label{sec:qMetric}

A more general spacetime is described by the q-metric shown in Eq~\eqref{q}. The aim of this section is to develop a neural network architecture capable of solving Einstein's field equations such that the solution adequately approximates the considered spacetime. Then, the network must solve the vacuum field equations ($R_{\mu\nu}=0$) for a general axially symmetric spacetime described by the metric
\begin{equation}\label{eq:axi}
    ds^2=-A(r,\theta)^2 dt^2 +B(r,\theta)^2 dr^2 +C(r,\theta)^2 d\theta^2+ D(r,\theta)^2 d\varphi^2\ ,
\end{equation}
where the metric functions $A_i=\{A,B,C,D\}$ must satisfy suitable boundary conditions so that the neural network approximation correctly reproduces the metric functions shown in Eq.~\eqref{q}.

The boundary conditions associated with the angular coordinate are periodic and are given by
\begin{equation}
    A_i(r,\theta=0)=A_i(r,\theta=\pi) \, .
\end{equation}
On the other hand, the boundary conditions associated with the radial coordinate require a more detailed treatment. In the case of Schwarzschild spacetime, the system requires only one boundary condition for the time component of the metric in the asymptotic limit $r\rightarrow \infty$, given by \eqref{eq.fSchw}. The requirement of a single boundary condition is not arbitrary, but rather a consequence of the structure of the field equations of general relativity. The freedom of coordinates allows a change of coordinates $r\rightarrow\Tilde{r}(r)$, through which a convenient relationship between $f(r)$ and $g(r)$ can be fixed. In particular, the Schwarzschild gauge $f(r)=1/g(r)$ fixes one of the metric functions and leaves only one independent function that requires a boundary condition. 
If a gauge is not explicitly imposed, the network may converge to a physically equivalent solution expressed in a different coordinate system. Then, the gauge can be explicitly imposed, or, as occurred in the previous section, emerge automatically during training.

\medskip

Therefore, the fundamental problem is to define appropriate boundary conditions that enable the neural network to approximate the q-metric. For this purpose, different scenarios are analyzed. 
The purpose of the following analysis is to determine how strongly the convergence of the PINN depends on the amount of boundary information supplied to the network.
First, boundary conditions are imposed on all four metric functions, corresponding to the strict case, which yields excellent results. However, when these conditions are progressively removed, the network does not converge. Consequently, we  propose an alternative solution strategy based on the domain decomposition technique, in which  the boundary conditions are imposed only on the outermost batch. In both cases, the solution obtained by the network, denoted by $A_i^{NN}$, must approximate the exact value $A_{i}^{exact}$ given by the q-metric shown in Eq~\eqref{q}.

\section{Strict Boundary Conditions}
\label{sec:StrictConditions}
An initial neural network is constructed with three hidden layers, each with 124 neurons. The network output corresponds to the set of functions $u_{i}(r,\theta)$, with $i=\{0,1,2,3\}$, which are related to the metric functions of the line element given in Eq~\eqref{eq:axi} by
\begin{align}
    A(r,\theta)^2&=\left(1-\frac{2m}{r}+\frac{u_0(r,\theta)}{r^2}\right)^{1+q}\label{eq:Total1}\\
    B(r,\theta)^2&=\left(1-\frac{2m}{r}+\frac{u_1(r,\theta)}{r^2}\right)^{-1-q}\left(1+\frac{m^2\sin^2\theta}{r^2-2mr}\right)^{-q(q+2)}\label{eq:Total2}\\
    C(r,\theta)^2&=\left(1-\frac{2m}{r}+\frac{u_2(r,\theta)}{r^2}\right)^{-q}\left(1+\frac{m^2\sin^2\theta}{r^2-2mr}\right)^{-q(q+2)}r^2\label{eq:Total3}\\
    D(r,\theta)^2&=\left(1-\frac{2m}{r}+\frac{u_3(r,\theta)}{r^2}\right)^{-q}r^2\sin^2(\theta) \label{eq:Total4}
\end{align}

These boundary conditions are called strict boundary conditions, where the parameters of the metric functions were taken as $q=1.0$ and $m=1.0$. The total computational domain is defined by the radial interval $r=(10m,120m)$ and the angular interval $\theta=(0,\pi)$ with $N_\theta=40$ collocation points respect to this angular coordinate. The radial domain is divided into six subintervals given by $[10,18.0,41.0,64.0,87.0,110.0,120]$, with the corresponding numbers of collocation points $[100,250,250,250,250,125]$.
The architecture of the neural network corresponds to five hidden layers, each with 128 neurons. The weight vector for the loss function was $[1.0,1.0,1.0,1.0,1.0,1.0]$ for each partial differential equation and the boundary conditions. We chose the {\tt Tanh} function as the activation function. The optimizer used was {\tt Adam}, with the usual configuration of parameters ({\tt betas:(0.9,0.99)}) and a learning rate of $1\times10^{-3}$, which was reduced by $10\%$ every 1000 iterations using an exponential decay scheme. The number of iterations of each subinterval were chosen as  $ [1001, 1001, 1001, 1001, 1001, 2501]$; however, training in each subinterval employed an early stop scheme, which  stops the training when the loss function reached an accuracy of $10^{-7}$ or when the training failed to improve after a specified number of steps. 

\begin{figure}[ht]
	\centering
	\begin{subfigure}{0.48\textwidth} 
		\includegraphics[width=\textwidth]{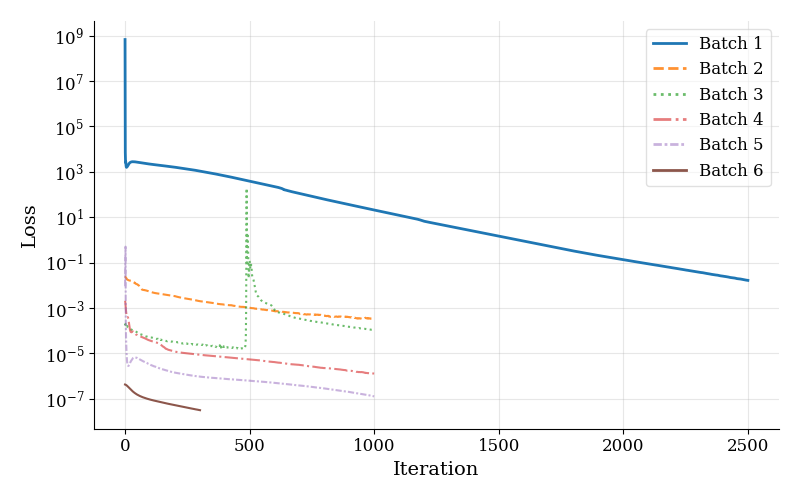}
		\caption{Total loss function.} 
	\end{subfigure}
	\vspace{1em} 
	\begin{subfigure}{0.48\textwidth} 
		\includegraphics[width=\textwidth]{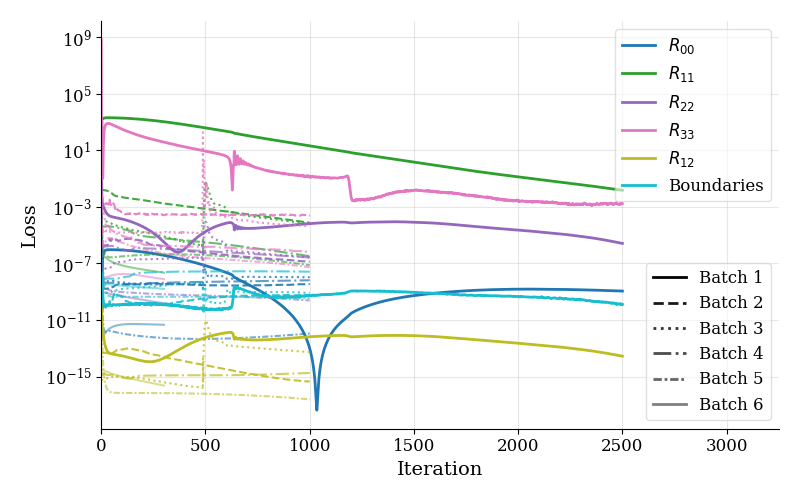}
		\caption{Loss function per equation.} 
	\end{subfigure}
	\caption{Loss function for strict boundary conditions, with the metric function parameters set to $m=1.0$ and $q=1.0$.} 
    \label{fig:LossStrictCase}
\end{figure}

The behavior of the total loss function, together with the individual contributions associated with each partial differential equation and each boundary equation, is presented in Figure \ref{fig:LossStrictCase}. After the training process, the optimization reaches a minimum loss function value of the order of $10^{-2}$ for the outermost subinterval, corresponding to batch 1. Subsequently, the inner subintervals easily achieve lower loss function values, even though they correspond to physical regions where the gravitational field is stronger. This is a consequence of the sequential training method, in which the weigths obtained during the first training are used to improve the subsequent ones.

\begin{figure}[H]
    \centering
    \includegraphics[width=0.95\linewidth]{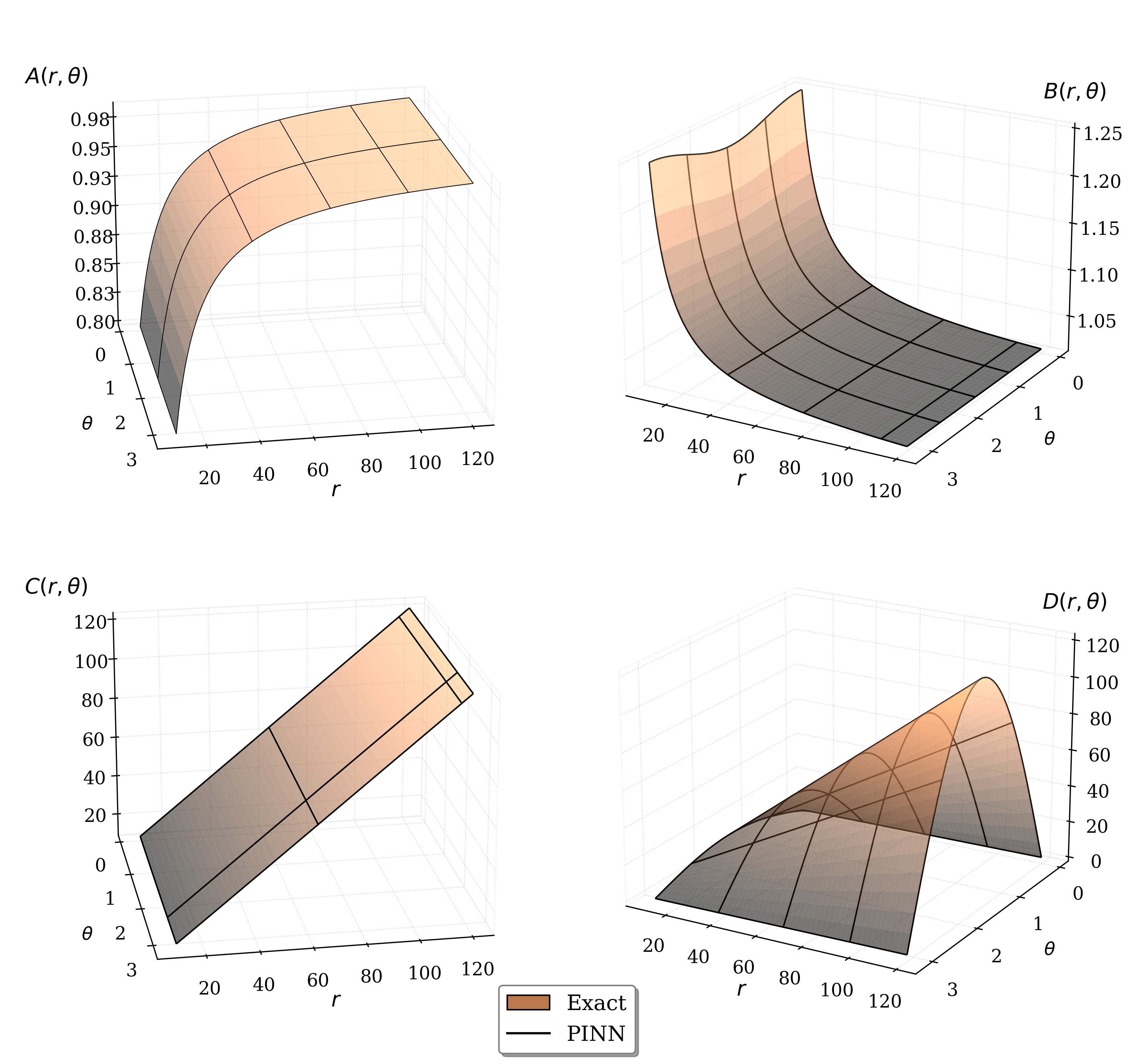}
    \caption{Three-dimensional visualizations of the metric functions and their respective approximations using the neural network, corresponding to the case of strict boundary conditions.}
    \label{fig:q1_T_w1_n128_1.0_3D_full}
\end{figure}

To analyze the accuracy of the  solution, the four exact metric functions are presented together with their corresponding approximations generated by the PINN for different values of the angular coordinate $\theta$. These comparisons are shown in Figures \ref{\labels{q1_T_w1_n128_1.0}{AthCutsVal}}, \ref{\labels{q1_T_w1_n128_1.0}{B_thCutsVal}}, \ref{\labels{q1_T_w1_n128_1.0}{C_thCutsVal}}, and \ref{\labels{q1_T_w1_n128_1.0}{D_thCutsVal}}. These results strongly suggest that the desired solution has been recovered, although this conclusion will be further supported by the complementary  analysis presented below. The three-dimensional plots presented in Fig. \ref{fig:q1_T_w1_n128_1.0_3D_full} also shows a good agreement between the exact solution and the PINN approximation. In particular, the metric function $B(r,\theta)$ exhibits a ridge-like structure in the inner region, arising from its angular dependence.  This feature is controlled by the parameter $q$ of the metric and is correctly reproduced by the neural network solution.

\figures{H}{0.95}{q1_T_w1_n128_1.0}{AthCutsVal}{Comparison between the exact metric function $A_{\text{exact}}$ and its corresponding neural network approximation $A_{\text{NN}}$, for different cuts of the angular coordinate $\theta$ in the case of strict boundary conditions.} 
\figures{H}{0.95}{q1_T_w1_n128_1.0}{B_thCutsVal}{Comparison between the exact metric function $B_{\text{exact}}$ and its corresponding neural network approximation $B_{\text{NN}}$, for different cuts of the angular coordinate $\theta$ in the case of strict boundary conditions.}
\figures{H}{0.95}{q1_T_w1_n128_1.0}{C_thCutsVal}{Comparison between the exact metric function $C_{\text{exact}}$ and its corresponding neural network approximation $C_{\text{NN}}$, for different cuts of the angular coordinate $\theta$ in the case of strict boundary conditions.}
\figures{H}{0.95}{q1_T_w1_n128_1.0}{D_thCutsVal}{Comparison between the exact metric function $D_{\text{exact}}$ and its corresponding neural network approximation $D_{\text{NN}}$, for different cuts of the angular coordinate $\theta$ in the case of strict boundary conditions.}

On the other hand, we present the local relative absolute error for each metric function in Fig.~\ref{\labels{q1_T_w1_n128_1.0}{RAE_ValidationDomain}}. It is observed that the network exhibits larger errors at small radii, close to the gravitational source, where the gradients of the metric functions are steeper. Nevertheless, these graphs show that the neural network reproduces the exact solution with errors of the order of $10^{-5}$ along the considered domain. Additionally, the relative $L_2$ error, equivalent to the relative RMSE for discretized data, is presented in Table~\ref{tab:RMSE_StrictCase}. Therefore, the global error of the metric functions is below $10^{-6}$. Hence, the network provides a highly accurate approximation of the solution. This behavior is consistent with the fact that, in the case considered here, the strict boundary conditions imply the trivial solution $u_{i_\text{NN}}=0$.


\begin{table}[H]
    \centering
    \vspace{0.8cm}
    \begin{tabular}{c c c c c}
        \toprule
        Metric Function & $A(r,\theta)$ & $B(r,\theta)$ & $C(r,\theta)$ & $D(r,\theta)$ \\
        \midrule
        RMSE & $\qquad 1.8\times10^{-6}\qquad $ & $6.0\times10^{-6}$ & $\qquad 5.1\times10^{-7}\qquad $ & $4.6\times10^{-7}$ \\
        \bottomrule
    \end{tabular}
    \caption{Global mean squared error (RMSE) values for each metric function, evaluated over an entire validation domain $r\in(10m,120m)$ and $\theta\in(0,\pi)$, under strict boundary conditions.}
    \label{tab:RMSE_StrictCase}
\end{table}

\figures{H}{1.0}{q1_T_w1_n128_1.0}{RAE_ValidationDomain}{Local absolute error of each metric function, computed as the absolute difference between the neural network prediction and the exact solution, $\epsilon_i=|A_{i}^{exact}-A_i^{NN}|$, under strict boundary conditions. }

\section{ Relaxed Boundary Conditions}
\label{sec:RelaxedConditions}

\begin{figure}[ht]
    \centering
    \includegraphics[width=1.0\linewidth]{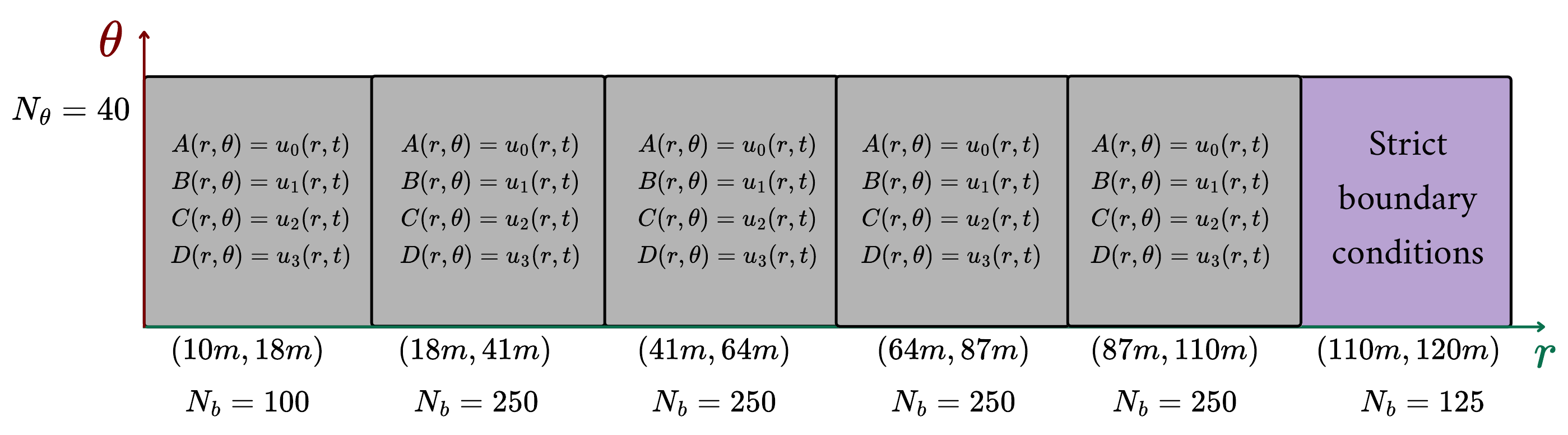}
    \caption{Schematic of the domain decomposition implemented in the neural network of Relaxed boundary conditions where the strict conditions are imposed only on the outer subdomain; the inner subdomains do not incorporate boundary conditions.}
    \label{fig:RelaxedScheme}
\end{figure}

To explore the robustness of PINNs, the next logical step is to relax the boundary conditions established in \eqref{eq:Total1}-\eqref{eq:Total4}.
The objective is to determine how much boundary information is actually required for a PINN to reconstruct a physically meaningful solution of Einstein's equations.
An initial strategy, based on gradually removing these conditions, does not produce satisfactory results. Consequently, an alternative approach is adopted: a neural network with domain decomposition, in which each subdomain (or {\it batch}) is trained progressively. Training process begins in the outermost subdomain, where strict boundary conditions are imposed; subsequently, the inner batches, without any boundary conditions, are incorporated into the training. The scheme of this decomposition is shown in Fig.~\ref{fig:RelaxedScheme}.

\vspace{5mm}

These boundary conditions are called relaxed boundary conditions, where, as in the previous case, the parameters of the metric functions were taken as $q=1.0$ and $m=1.0$. The total computational domain is defined by the radial interval $r=(10m,120m)$ and the angular interval $\theta=(0,\pi)$, with $N_\theta=40$ collocation points with respect to this angular coordinate. Nevertheless, the six subdomains of the radial domain $[10.0,18.0,41.0,64.0,87.0,110.0,120.0]$, with the corresponding numbers of collocation points $[100,250,250,250,250,125]$ now take relevance, as only in the outermost case, the strict boundary conditions were established.
The architecture of the neural network has the same parameters as in the previous case: the network has five hidden layers, each with 128 neurons; the vector of weights for the loss function was $[1.0,1.0,1.0,1.0,1.0,1.0]$; the activation function was {\tt Tanh}; the optimizer used was {\tt Adam}, with the usual configuration of parameters ({\tt betas:(0.9,0.99)}) and a learning rate of $1\times10^{-3}$ with an exponential decay scheme. The number of iterations of each subinterval was $ [1001, 1001, 1001, 1001, 1001, 2501]$, with the training in each subinterval employing the same early-stopping scheme as before.

\medskip

\begin{figure}[H]
	\centering
	\begin{subfigure}{0.48\textwidth} 
		\includegraphics[width=\textwidth]{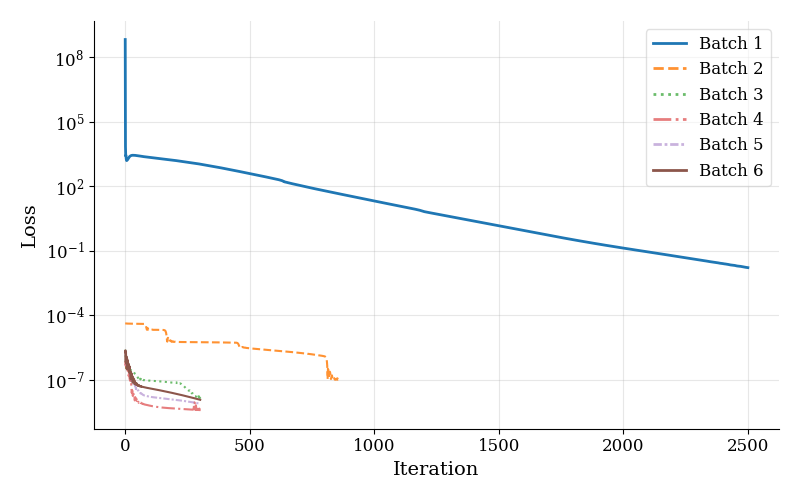}
		\caption{Total loss function.} 
	\end{subfigure}
	\vspace{1em} 
	\begin{subfigure}{0.48\textwidth} 
		\includegraphics[width=\textwidth]{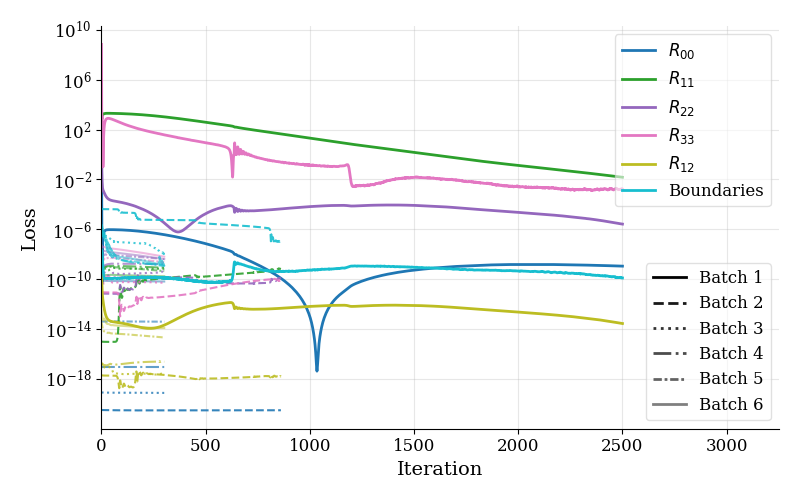}
		\caption{Loss function per equation.} 
	\end{subfigure}
	\caption{Loss function for relaxed boundary condition, with the metric function parameters set to $m=1.0$ and $q=1.0$.} 
    \label{fig:LossRelaxed}
\end{figure}

The total loss function, together with the individual contributions associated with each partial differential equation and each boundary equation, is presented in Fig.~\ref{fig:LossRelaxed}. After the training iterations, the optimization process reaches a minimum loss function value of the order of $10^{-2}$ for the outermost subinterval. In general, as observed in the previous section, after successfully training the outermost batch, corresponding to batch 1, the remaining subdomains exhibit significantly lower loss function values, even though they correspond to physical regions where the gravitational field is stronger.

\medskip

The four exact metric functions are presented together with their corresponding approximations generated by the PINN for different values of the angular coordinate $\theta$. These results are shown in Figures \ref{\labels{q1_F_w1_n128_1.0}{AthCutsVal}}, \ref{\labels{q1_F_w1_n128_1.0}{B_thCutsVal}}, \ref{\labels{q1_F_w1_n128_1.0}{C_thCutsVal}}, and \ref{\labels{q1_F_w1_n128_1.0}{D_thCutsVal}}. Again, the results strongly suggest that the desired solution has been recovered, but this conclusion must be supported by additional error analysis, presented below. The three-dimensional plots shown in Fig.~\ref{\labels{q1_F_w1_n128_1.0}{3D_full}} also show good agreement between the exact solution and the PINN approximation. In particular, the ridge-like structure in the inner region of the metric function $B(r,\theta)$ is again correctly reproduced by the neural network approximation.

\figures{H}{1}{q1_F_w1_n128_1.0}{AthCutsVal}{Comparison between the exact metric function $A_{\text{exact}}$ and its corresponding neural network approximation $A_{\text{NN}}$, for different cuts of the angular coordinate $\theta$ in the case of relaxed boundary conditions.} 
\figures{H}{1}{q1_F_w1_n128_1.0}{B_thCutsVal}{Comparison between the exact metric function $B_{\text{exact}}$ and its corresponding neural network approximation $B_{\text{NN}}$, for different cuts of the angular coordinate $\theta$ in the case of relaxed boundary conditions.}
\figures{H}{1}{q1_F_w1_n128_1.0}{C_thCutsVal}{Comparison between the exact metric function $C_{\text{exact}}$ and its corresponding neural network approximation $C_{\text{NN}}$, for different cuts of the angular coordinate $\theta$ in the case of relaxed boundary conditions.}
\figures{H}{1}{q1_F_w1_n128_1.0}{D_thCutsVal}{Comparison between the exact metric function $D_{\text{exact}}$ and its corresponding neural network approximation $D_{\text{NN}}$, for different cuts of the angular coordinate $\theta$ in the case of relaxed boundary conditions.}

\figures{H}{1}{q1_F_w1_n128_1.0}{3D_full}{Three-dimensional visualizations of the metric functions and their respective approximations using the physics-informed neural network, corresponding to the case of relaxed boundary conditions.}

\medskip

Regarding the error analysis, we present the local relative absolute error for each metric function in the case of relaxed boundary conditions in Fig.~\ref{\labels{q1_F_w1_n128_1.0}{RAE_ValidationDomain}}. As in the case of strict conditions, it is observed that the network exhibits larger errors at small radii, close to the gravitational source, where the gradients of the metric functions are steeper. These graphs show that the neural network reproduces the exact solution with errors below the order of $10^{-3}$ over the considered domain. Additionally, the relative $L_2$ error, or relative RMSE, is presented in Table~\ref{tab:RMSE_RelaxedCase}. Therefore, the global error of the metric functions is below $10^{-4}$. Hence, the network provides an accurate approximation of the solution even when the boundary conditions are relaxed.

\medskip

\figures{H}{1.0}{q1_F_w1_n128_1.0}{RAE_ValidationDomain}{Local absolute error of each metric function, computed as the absolute difference between the neural network prediction and the exact solution, $\epsilon_i=|A_{i}^{exact}-A_i^{NN}|$, under relaxed boundary conditions.}

\begin{table}[H]
    \centering
    \vspace{0.8cm}
    \begin{tabular}{c c c c c}
        \toprule
        Metric Function & $A(r,\theta)$ & $B(r,\theta)$ & $C(r,\theta)$ & $D(r,\theta)$ \\
        \midrule
        RMSE & $\qquad 5.5\times10^{-6}\qquad $ & $1.0\times10^{-4}$ & $\qquad 2.5\times10^{-5}\qquad $ & $5.5\times10^{-5}$ \\
        \bottomrule
    \end{tabular}
    \caption{Global mean squared error (RMSE) values for each metric function, evaluated over an entire validation domain $r\in(10m,120m)$ and $\theta\in(0,\pi)$, under relaxed boundary conditions.}
    \label{tab:RMSE_RelaxedCase}
\end{table}

\newpage
\subsection{Minkowski Limit ($q=-1$)}
\label{sec:qMetricMink}

The q metric includes the case of flat Minkowski spacetime when the quadrupole parameter takes the value $q=-1$. However, in this limit the metric takes the following form
\begin{align}
    ds^2=&-dt^2+\left(1-\frac{2m}{r}\right)\nonumber\\
    &\times \left[\left(1+\frac{m^2\sin^2\theta}{r^2-2mr}\right)\left[r^2d\theta^2+\frac{dr^2}{1-2m/r}\right]+r^2\sin^2\theta\ d\varphi^2 \right]\ .\label{qMink}
\end{align}
Therefore, although the q-metric formally includes flat spacetime as a limiting case, it is not immediately clear from the explicit form of the metric in Eq.~\eqref{qMink} that the corresponding spacetime is indeed Minkowski spacetime. This is, in fact, true, as can be demonstrated by calculating the Riemann curvature tensor, which vanishes, $R_{\mu\nu\alpha\beta}=0$, for all indices when $q=-1$. Similarly, the multipole moments presented in Eq.~\eqref{momentosMul} are proportional to $(q+1)$ and therefore vanish in this limit. Consequently, the spacetime lacks both multipolar structure and curvature, corresponding geometrically to a flat spacetime.

\medskip

From a computational point of view, this case is particularly relevant, since obtaining a flat spacetime from an approximate solution of the system is not straightforward. Physics-informed neural networks are not restricted to reproducing a specific choice of coordinates and  instead converge to equivalent representations related by coordinate transformations. Therefore, even when the spacetime is flat, the solution learned by the network does not necessarily coincide with the standard form of the Minkowski metric. This makes it essential to draw physical conclusions from invariant or physically relevant quantities, such as scalars constructed from the Riemann tensor, rather than relying solely on the explicit form of the metric components.

Consequently, the study of the case $q=-1$ is fundamental from both geometric and numerical perspectives, since the line element does not adopt the standard form of flat spacetime in spherical coordinates. Therefore, flatness cannot be inferred by simple inspection, so a more in-depth analysis based on geometrically relevant quantities is required.

To begin this analysis, we use the same model considered in the previous case, composed of five hidden layers, each with 128 neurons and with the same hyperparameters (such as the activation function and the optimization parameters). The training process of the network of this limiting case yields the results shown in the figures below.

The loss function is presented in Fig.~\ref{fig:LossRelaxedqn1}, both for each batch and for each differential or boundary equation. The outermost subinterval reaches a minimum loss value of the order of $10^{-4}$ during the optimization process. As in previous sections, after successfully training this batch, the remaining subdomains exhibit significantly lower loss function values. In comparison with the previous case, in this limiting case the training process exhibits improved behavior, as the loss function reaches lower values within the same number of iterations.

\medskip

\begin{figure}[H]
	\centering
	\begin{subfigure}{0.49\textwidth} 
		\includegraphics[width=\textwidth]{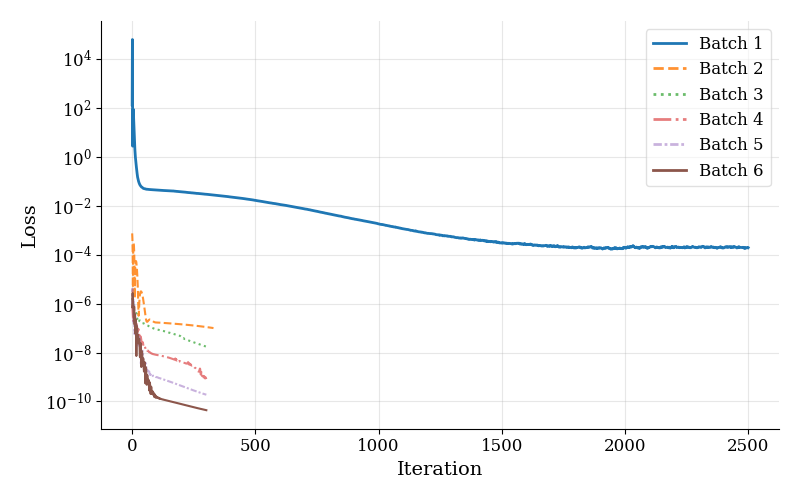}
		\caption{Total loss function.} 
	\end{subfigure}
	\vspace{1em} 
	\begin{subfigure}{0.49\textwidth} 
		\includegraphics[width=\textwidth]{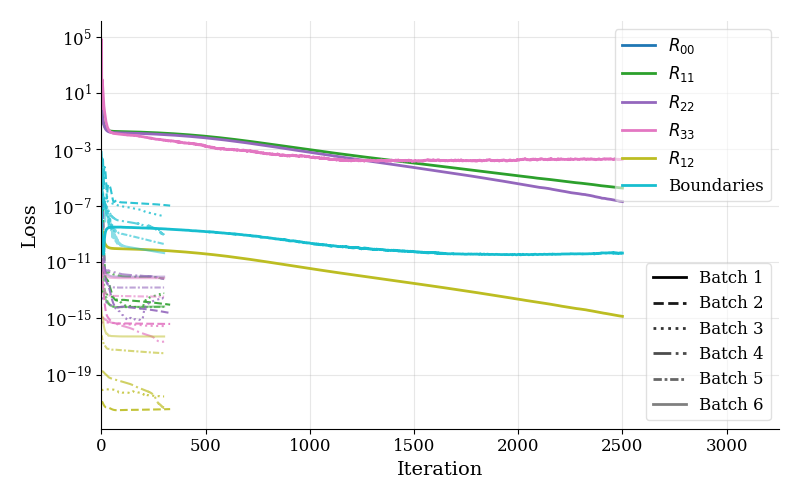}
		\caption{Loss function per equation.} 
	\end{subfigure}
	\caption{Loss function for the relaxed boundary condition, with the metric function parameters set to $m=1.0$ and $q=-1.0$, corresponding to the Minkowski limit.} 
    \label{fig:LossRelaxedqn1}    
\end{figure}

As in the previous case, the four exact metric functions are presented together with their corresponding approximations generated by the PINN for different values of the angular coordinate $\theta$. These results are shown in Figures \ref{\labels{qn1_F_w1_n128_-1.0}{AthCutsVal}}, \ref{\labels{qn1_F_w1_n128_-1.0}{B_thCutsVal}}, \ref{\labels{qn1_F_w1_n128_-1.0}{C_thCutsVal}}, and \ref{\labels{qn1_F_w1_n128_-1.0}{D_thCutsVal}}. Once again, the results suggest strong agreement with the desired solution, a conclusion that will be corroborated by the error analysis presented below. The three-dimensional plots shown in Fig.~\ref{\labels{qn1_F_w1_n128_-1.0}{3D_full}} further confirm the close correspondence between the exact solution and the PINN approximation. Additionally, the ridge-like structure of the metric function $B(r,\theta)$ disappears in both the exact solution and the neural network approximation, as expected for a configuration with a quadrupole parameter $q=-1$.

\figures{H}{0.95}{qn1_F_w1_n128_-1.0}{AthCutsVal}{Comparison between the exact metric function $A_{\text{exact}}$ and its corresponding neural network approximation $A_{\text{NN}}$, for different cuts of the angular coordinate $\theta$ in the case of relaxed boundary conditions, with the parameter set to $m=1.0$ and $q=-1.0$, corresponding to the Minkowski limit.} 
\figures{H}{0.95}{qn1_F_w1_n128_-1.0}{B_thCutsVal}{Comparison between the exact metric function $B_{\text{exact}}$ and its corresponding neural network approximation $B_{\text{NN}}$, for different cuts of the angular coordinate $\theta$ in the case of relaxed boundary conditions, with the parameter set to $m=1.0$ and $q=-1.0$, corresponding to the Minkowski limit.}
\figures{H}{0.95}{qn1_F_w1_n128_-1.0}{C_thCutsVal}{Comparison between the exact metric function $C_{\text{exact}}$ and its corresponding neural network approximation $C_{\text{NN}}$, for different cuts of the angular coordinate $\theta$ in the case of relaxed boundary conditions, with the parameter set to $m=1.0$ and $q=-1.0$, corresponding to the Minkowski limit.}
\figures{H}{0.95}{qn1_F_w1_n128_-1.0}{D_thCutsVal}{Comparison between the exact metric function $D_{\text{exact}}$ and its corresponding neural network approximation $D_{\text{NN}}$, for different cuts of the angular coordinate $\theta$ in the case of relaxed boundary conditions, with the parameter set to $m=1.0$ and $q=-1.0$, corresponding to the Minkowski limit.}

\figures{H}{0.95}{qn1_F_w1_n128_-1.0}{3D_full}{Three-dimensional visualizations of the metric functions and their respective approximations using the physics-informed neural network, corresponding to the case of relaxed boundary conditions and the Minkowski limit ($q=-1$).}

Regarding the error analysis, the local relative absolute error for each metric function under relaxed boundary conditions, with the quadrupole parameter set to $q=-1$, is shown in Fig.~\ref{\labels{q1_F_w1_n128_1.0}{RAE_ValidationDomain}}. As in the previous cases, the neural network approximation exhibits larger errors at small radii for the metric functions $C(r,\theta)$ and $D(r,\theta)$ below $10^{-4}$ over the considered domain. In contrast, the approximation of the metric function $A(r,\theta)$ agrees exactly with the exact solution throughout the considered domain. Similarly, the relative error for the metric function $B(r,\theta)$ is nearly zero over most of the domain, with only a few localized regions exhibiting errors of the order of $10^{-7}$. Additionally, the relative $L_2$ error, or relative RMSE, is presented in Table~\ref{tab:RMSE_RelaxedCase}. Therefore, this global relative error of the metric functions is below $10^{-6}$. Hence, the network provides an even more accurate approximation of the solution in this limiting case, characterized by the absence of gravitational curvature. Interestingly, the approximation of the metric function $A(r,\theta)$ coincides exactly with the exact solution throughout the considered domain. Since the exact solution is simply $A(r,\theta)=1$, the network successfully reproduces this constant solution throughout the computational domain.

\figures{H}{1.0}{qn1_F_w1_n128_-1.0}{RAE_ValidationDomain}{Local absolute error of each metric function, computed as the absolute difference between the neural network prediction and the exact solution, $\epsilon_i=|A_{i}^{exact}-A_i^{NN}|$, under relaxed boundary conditions and in the Minkowski limit ($q=-1$).}

\begin{table}[H]
    \centering
    \vspace{0.8cm}
    \begin{tabular}{c c c c c}
        \toprule
        Metric Function & $A(r,\theta)$ & $B(r,\theta)$ & $C(r,\theta)$ & $D(r,\theta)$ \\
        \midrule
        RMSE & $\qquad 0.0\qquad $ & $4.3\times10^{-9}$ & $\qquad 6.3\times10^{-6}\qquad $ & $3.0\times10^{-6}$ \\
        \bottomrule
    \end{tabular}
    \caption{Global mean squared error (RMSE) values for each metric function, evaluated over an entire validation domain $r\in(10m,120m)$ and $\theta\in(0,\pi)$, under relaxed boundary conditions and in the Minkowski limit ($q=-1$).}
    \label{tab:RMSE_RelaxedCaseqn1}
\end{table}

As mentioned at the beginning of this section, it is essential to establish, through numerical evidence, that the spacetime obtained by the neural network effectively corresponds to the flat spacetime. To this end, a module was developed and implemented to compute the components of the Riemann curvature tensor for a general static and axially symmetric spacetime, described by Eq.~\eqref{eq:axi}. The resulting error metrics for these non-trivial components are presented in Table~\ref{tab:Riemannqn1}. The RMSE and $L_2$ values remain below $10^{-4}$ and $10^{-2}$, respectively, while the $L_\infty$ values are of the order of $10^{-4}$ across all components. The larger errors observed in the Riemann tensor components, compared with those obtained for the metric functions, can be explained by the fact that their computation involves second-order derivatives of the metric functions, which can amplify approximation errors. In particular, the $L_2$ norm accumulates contributions from all points in the domain, whereas the $L_\infty$ norm is determined by the largest individual error. Therefore, the first is generally larger than the second. We focus on the $L_\infty$ norm, as it directly measures the maximum pointwise error. 

\medskip

In this way, since the exact Riemann components vanish identically for $q=-1$, the nonzero values obtained from the neural network approximation can be interpreted as residual numerical curvature introduced by the approximation error, quantified by an $L_\infty$ norm and of the order of $10^{-4}$. In other words, our computed PINN Riemann tensor effectively measures the residual curvature introduced by the neural network approximation. Despite these errors, their small magnitude throughout the computational domain provides numerical evidence that the spacetime represented by the neural network is effectively flat.

These results indicate that Einstein's equations alone are generally insufficient to guide the optimization toward the desired physical solution. The amount and distribution of boundary information play a fundamental role in selecting the physically relevant solution among the many functions explored during the training process. Domain decomposition provides an effective mechanism for propagating this information from well-constrained regions into less constrained parts of the computational domain.

\begin{table}[H]
    \centering
    \begin{tabular}{c c  c  c}
    \toprule
        Riemann Component & RMSE & $L_2$ & $L_\infty$ \\
        \midrule
         $R_{trtr}$ & $0.0$ & $0.0$&$0.0$\\  
         $R_{trt\theta}$ & $0.0$ & $0.0$&$0.0$\\  
         $R_{t\theta t\theta}$ & $0.0$ & $0.0$&$0.0$\\  
         $R_{t\varphi t\varphi}$ & $0.0$ & $0.0$&$0.0$\\  
         $R_{r\theta r\theta }$ & $\quad5.7\times 10^{-5}\quad$ & $\quad8.0\times10^{-3}\quad$&$\quad3.5\times10^{-4}\quad$\\  
         $R_{r\varphi r\varphi }$ & $3.9\times 10^{-5}$ & $5.5\times10^{-3}$&$4.0\times10^{-4}$\\  
         $R_{r\varphi \theta\varphi }$ & $1.6\times 10^{-4}$ & $2.2\times10^{-2}$&$4.3\times10^{-4}$\\  
         $R_{\theta\varphi \theta\varphi }$ & $2.4\times 10^{-5}$ & $3.5\times10^{-3}$&$2.6\times10^{-4}$\\  
        \bottomrule
    \end{tabular}
    \caption{Global error metrics for the non-trivial Riemann tensor components computed from the PINN approximation in the $q=-1$ limiting case, for the line element given by Eq.~\eqref{eq:axi}. The RMSE, $L_2$ norm, and $L_\infty$ norm quantify the deviation of each component from its exact value of zero.}
    \label{tab:Riemannqn1}
\end{table}


\section{Validation Methods}
\label{sec:VM}

In traditional numerical methods, the convergence order is commonly evaluated by analyzing how the numerical error scales with the spatial resolution, according to $E(h)\sim h^p$, where $h$ is the spatial discretization scale and $p$ is the order of convergence of the numerical method. 
However, this approach is not directly applicable to PINNs, which do not require a conventional mesh-based discretization of the governing equations and whose training process is based on minimizing the equation residuals at selected collocation points. 
These points are used to evaluate the residuals, while the neural network gives a continuous approximation solution across the computational domain. Besides, the optimization procedure itself may introduce stochastic effects.

Therefore, the accuracy of a PINN solution is commonly examined by comparing it either with a known analytical solution or with a reference solution obtained using an established numerical method. In the cases analyzed in the previous sections, the accuracy of the PINN solutions was evaluated using local and global error metrics with respect to the known analytical solution. For example, the relative RMSE provides a dimensionless measure of the prediction error relative to the corresponding solution variable. When the exact solution vanishes, as in the case of the Ricci and Riemann tensor components for the $q=-1$ limiting case, the absolute RMSE is used instead, since a relative error is not defined. Unlike the relative RMSE, the absolute RMSE provides a direct measure of the prediction error in the same units as the corresponding solution variable. These errors were computed on an independent validation set that covers the same physical domain as the training set but uses a different distribution of points, with $N_{\theta_{\mathrm{val}}}=40$ and $N_{r_{\mathrm{val}}}=300$ validation points in the angular and radial directions, respectively. An independent validation set allows the accuracy of the learned solution to be evaluated at points not used during training, providing a more reliable evaluation of the network's performance across the computational domain.


\medskip

\figures{H}{1.0}{q1_T_w1_n128_1.0}{RAE_Ricci_ValidationDomain}{Absolute error of the non-trivial Ricci tensor components computed from the PINN approximation under strict boundary conditions, with the metric parameters set to $m=1.0$ and $q=1.0$ (section \ref{sec:StrictConditions}).}

\medskip

\figures{H}{1.0}{q1_F_w1_n128_1.0}{RAE_Ricci_ValidationDomain}{Absolute error of the non-trivial Ricci tensor components computed from the PINN approximation under relaxed boundary conditions, with the metric parameters set to $m=1.0$ and $q=1.0$ (section \ref{sec:RelaxedConditions}).}

\medskip

In this context, we implemented an additional validation method by computing the Ricci tensor over the same validation domain. Consequently, the absolute errors of the non-trivial Ricci tensor components for the cases presented in Section~\ref{sec:StrictConditions} and~\ref{sec:RelaxedConditions} are shown in Figures~\ref{\labels{q1_T_w1_n128_1.0}{RAE_Ricci_ValidationDomain}} and~\ref{\labels{q1_F_w1_n128_1.0}{RAE_Ricci_ValidationDomain}}, respectively. The resulting RMSE is below $10^{-5}$ for the strict boundary condition case and below $10^{-4}$ for the relaxed boundary condition case, providing an additional validation of the obtained solution independent of the analytical solution. 

\medskip

On the other hand, as discussed in Section~\ref{sec:qMetricMink}, the network may converge to an equivalent coordinate form of the solution. Therefore, it is also important to validate the solution using geometrically meaningful scalar quantities. To this end, we computed the Kretschmann scalar from the PINN approximation for the cases presented in Section~\ref{sec:StrictConditions} and~\ref{sec:RelaxedConditions}. The corresponding results are shown in Figures~\ref{\labels{q1_T_w1_n128_1.0}{K_ValidationDomain}} and~\ref{\labels{q1_F_w1_n128_1.0}{K_ValidationDomain}}, respectively. The Kretschmann scalar shows good agreement between the exact and PINN-approximated values, with an RMSE of the order of $10^{-4}$ for the first case and $10^{-2}$ for the second. In both cases, the Ricci tensor and the  Kretschmann scalar provide geometrically meaningful quantities that allow us to validate the PINN solution through quantities involving up to second-order derivatives of the metric functions, providing a more rigorous approach to test the final obtained solution.

\figures{H}{1.0}{q1_T_w1_n128_1.0}{K_ValidationDomain}{Kretschmann scalar computed from the PINN approximation under strict boundary conditions, with the metric parameters set to $m=1.0$ and $q=1.0$ (Section~\ref{sec:StrictConditions}). The right panel shows the three-dimensional comparison between the exact value and PINN approximated solution, while the left panel shows the corresponding relative absolute error.}

\figures{H}{1.0}{q1_F_w1_n128_1.0}{K_ValidationDomain}{Kretschmann scalar computed from the PINN approximation under relaxed boundary conditions, with the metric parameters set to $m=1.0$ and $q=1.0$ (Section~\ref{sec:RelaxedConditions}). The right panel shows the three-dimensional comparison between the exact value and PINN approximated solution, while the left panel shows the corresponding relative absolute error.}

Additionally, it is important to emphasize that, in PINN, a reduction in the loss function does not necessarily guarantee convergence to the physical solution. For example, in the case of relaxed boundary conditions  presented in Section~\ref{sec:RelaxedConditions}, the loss function reaches values of the order of $10^{-2}$, whereas the RMSE is below $10^{-4}$. This difference illustrates that the loss function and the RMSE quantify different aspects of the solution: the first measures the residuals of the governing equations and boundary conditions, while the latter directly measures the deviation from the known analytical solution. A poor balance among the different contributions to the loss function, the dominance of specific terms, or under-sampled regions of the computational domain may lead to solutions that satisfy some equations or conditions more accurately than others. In the context of GR, the coupled and nonlinear structure of the partial differential equation from the vacuum Einstein equations can further complicate this issue, as some equations may be satisfied to a higher degree of accuracy than others. This behavior can be mitigated by appropriately weighting the different contributions to the loss function. Furthermore, because the loss function is constructed from the Einstein vacuum equation, which involve second-order derivatives of the metric function, small errors in the metric approximation can result in larger residuals in the governing equations. Consequently, a complete validation of a PINN solution should consider not only the total loss but also local residuals, physical constraints, and out-of-training validation points, such as those given by an independent validation set. 

\medskip

In general, a significantly low value of the loss function is a strong indication that the network has learned a good approximation of the solution, although it is not a guarantee of physical accuracy. In the case discussed above, the loss function also exhibits a globally monotonic decreasing behavior, suggesting that further training may lead to additional improvements in the solution. However, the main purpose of the presented analysis is not to obtain an arbitrary precise numerical solution, but rather to investigate how additional validation mechanisms can be employed in general relativity to assess the physical consistency of PINN solutions beyond the total function alone.




\newpage
\section{Conclusions}
\label{sec:con}

In this work, we investigated a novel tool in the context of General Relativity theory: Physics-Informed Neural Networks (PINNs). To this end, we developed two PINNs in which different boundary conditions were implemented. The first network approximates the Schwarzschild metric by solving the field equations that describe an empty, spherically symmetric spacetime. The architecture employed consists of three hidden layers with 64 neurons each, and we used the domain decomposition technique for the radial domain $r\in(10m,100m)$. The network achieves a minimum loss on the order of $10^{-7}$, in the outermost domain. The domain decomposition method significantly improves accuracy over the radial domain. The resulting RMSE values are of the order of $10^{-3}$ and $10^{-8}$ for the temporal and radial component of the Schwarzschild metric, respectively.

\medskip

The second model developed in this work approximates the q-metric; that is, it solves for the metric functions that describe an empty, axially symmetric spacetime. The model was investigated under two types of boundary conditions: strict boundary conditions, presented in Section~\ref{sec:StrictConditions}, for which the network must learn the trivial solution, and relaxed boundary conditions, presented in Section \ref{sec:RelaxedConditions}, for which boundary conditions are imposed only on the outermost radial subdomain. In both cases, the deep neural networks consist of five hidden layers with 128 neurons per layer, and the metric parameters are set to $q=1.0$ and $m=1.0$. In the case of the strict boundary conditions, the loss function reaches a value of the order of $10^{-2}$ in the outermost subdomain. A complete set of plots was created to study the accuracy of the obtained solution. It is observed that, for all metric functions, the relative RMSE remains below $10^{-6}$, reaching values on the order of $10^{-7}$ for the metric functions $C(r, \theta)$ and $D(r, \theta)$. Consequently, the constructed neural network reproduces the exact solution with a good accuracy throughout the considered domain.

To relax the boundary conditions established in the first case, we use the domain decomposition method, in which the strict boundary conditions are applied only to the outermost subdomain. The architecture used has the same hyperparameters as in the first case. The obtained relative RMSE is of the order of $10^{-4}$ for the metric function $B(r,\theta)$, $10^{-5}$ for the metric functions $C(r,\theta)$ and $D(r,\theta)$, and $10^{-6}$ for the metric function $A(r,\theta)$. Therefore, as in the first case, the constructed neural network reproduces the exact solution with good accuracy throughout the considered domain, even under more general boundary conditions. 
These results provide practical guidelines for the successful application of Physics-Informed Neural Networks to gravitational configurations for which analytical solutions are unavailable.

We also investigated the Minkowski limiting case in Subsection~\ref{sec:qMetricMink}, where the quadrupole parameter is set to $q=-1$, corresponding to a flat spacetime. In this scenario, the relative RMSE for the metric functions $C(r,\theta)$ and $D(r,\theta)$ is of the order of $10^{-6}$, while that for $B(r,\theta)$ is of the order of $10^{-9}$. Meanwhile, the metric function $A(r,\theta)$ coincides exactly with its analytical value, $A(r,\theta)=1.0$. Thus, the PINNs accurately reproduce the solution for $q=-1$, with even better accuracy than in the previous cases, as the quadrupolar contribution is not present in this limit. This case is particularly relevant because, although the spacetime is geometrically flat, the line element does not adopt the standard form of the Minkowski metric in spherical coordinates. 
For this reason, the Riemann curvature tensor for a static and axially symmetric spacetime is calculated and evaluated using the solution obtained by the network. The $L_\infty$ norms of the non-trivial Riemann tensor components are found to be of the order of $10^{-4}$. Since the exact Riemann tensor components vanish identically for $q=-1$, these nonzero values can be interpreted as residual numerical curvature introduced by the approximation error. In conclusion, despite these errors, their small magnitude throughout the computational domain provides numerical evidence that the spacetime learned by the neural network is effectively flat.


\medskip

These results show that Physics-Informed Neural Networks are not restricted to reproducing the metric in a particular coordinate representation of a certain coordinate system, but can instead converge to equivalent forms of the solution related by coordinate transformations. 
This makes it essential to draw physical conclusions from invariant or physically meaningful quantities, such as scalars constructed from the Riemann tensor, rather than basing the analysis entirely on the explicit form of the metric components. 

Consequently, we presented the Ricci tensor components and the Kretschmann scalar for both the strict and relaxed boundaries conditions as additional validation methods for assessing the physical consistency of the obtained solutions.




\acknowledgments

This work  was partially supported by  PAPIIT-DGAPA-UNAM, grant No. 108225, and by CONAHCYT,  grant No. CBF-2025-I-253.


\bibliography{0bibliografy}

@article{stefanou2023pulsar,
  title={Solving the pulsar equation using physics-informed neural networks},
  author={Stefanou, Petros and Urb{\'a}n, Jorge F and Pons, Jos{\'e} A},
  journal={Monthly Notices of the Royal Astronomical Society},
  volume={526},
  number={1},
  pages={1504--1511},
  year={2023},
  publisher={Oxford University Press}
}

@article{QNSchw,
  title={Calculating Quasi-Normal Modes of Schwarzschild Black Holes with Physics Informed Neural Networks},
  author={Patel, Nirmal and Aykutalp, Aycin and Laguna, Pablo},
  journal={arXiv preprint arXiv:2401.01440},
  year={2024}
}

@article{li2023solvingSchw,
  title={Solving Einstein equations using deep learning},
  author={Li, Zhi-Han and Li, Chen-Qi and Pang, Long-Gang},
  journal={arXiv preprint arXiv:2309.07397},
  year={2023}
}

@article{qMetric_24,
  title={Topology of some spheroidal metrics},
  author={Zipoy, David M},
  journal={Journal of Mathematical Physics},
  volume={7},
  number={6},
  pages={1137--1143},
  year={1966},
  publisher={American Institute of Physics}
}

@article{qMetric_25,
  title={Static axially symmetric gravitational fields},
  author={Voorhees, BH},
  journal={Physical Review D},
  volume={2},
  number={10},
  pages={2119},
  year={1970},
  publisher={APS}
}

@article{qMetric_21,
  title={Monopole-quadrupole static axisymmetric solutions of Einstein field equations},
  author={Hern{\'a}ndez-Pastora, JL and Mart{\'\i}n, J},
  journal={General Relativity and Gravitation},
  volume={26},
  number={9},
  pages={877--907},
  year={1994},
  publisher={Springer}
}

@article{qMetric_23, 
  title={Mass quadrupole as a source of naked singularities},
  author={Quevedo, Hernando},
  journal={International Journal of Modern Physics D},
  volume={20},
  number={10},
  pages={1779--1787},
  year={2011},
  publisher={World Scientific}
}

@article{qMetric_26,
  title={Physical properties of the sources of the gamma metric},
  author={Malafarina, D},
  journal={Dynamics and Thermodynamics of Blackholes and Naked Singularities},
  pages={20},
  year={2004}
}

@article{qMetric_28, 
  title={Motion of test particles in the field of a naked singularity},
  author={Boshkayev, K and Gasperin, E and Guti{\'e}rrez-Pi{\~n}eres, AC and Quevedo, H and Toktarbay, S},
  journal={Physical Review D},
  volume={93},
  number={2},
  pages={024024},
  year={2016},
  publisher={APS}
}

@incollection{Quevedo2017Quadrupolar,
  author    = {Hernando Quevedo},
  title     = {Quadrupolar Metrics},
  booktitle = {Neutron Stars: Physics Properties and Dynamics},
  editor    = {Nurgali Takibayev and Kuantay Boshkayev},
  publisher = {Nova Science Publishers},
  series    = {Physics Research and Technology},
  year      = {2017},
  isbn      = {978-1-53610-525-4},
  eprint    = {1606.09361},
  archivePrefix = {arXiv}
}

@article{QuevedoPedro,
  year = {2018},
  month = may,
  publisher = {The Royal Society},
  volume = {5},
  number = {5},
  pages = {170826},
  author = {Francisco Frutos-Alfaro and Hernando Quevedo and Pedro A. Sanchez},
  title = {Comparison of vacuum static quadrupolar metrics},
  journal = {Royal Society Open Science}
}

@book{alcubierre2008introduction,
  title={Introduction to 3+ 1 numerical relativity},
  author={Alcubierre, Miguel},
  volume={140},
  year={2008},
  publisher={OUP Oxford}
}

@article{geroch1970multipole2,
  title={Multipole moments. II. Curved space},
  author={Geroch, Robert},
  journal={Journal of Mathematical Physics},
  volume={11},
  number={8},
  pages={2580--2588},
  year={1970},
  publisher={American Institute of Physics}
}

@article{geroch1970multipole1,
  title={Multipole moments. I. Flat space},
  author={Geroch, Robert},
  journal={Journal of Mathematical Physics},
  volume={11},
  number={6},
  pages={1955--1961},
  year={1970},
  publisher={American Institute of Physics}
}

@article{Raissi2019PINN,
  author = {Raissi, Maziar and Perdikaris, Paris and Karniadakis, George E.},
  title = {Physics-informed neural networks: A deep learning framework for solving forward and inverse problems involving nonlinear partial differential equations},
  journal = {Journal of Computational Physics},
  volume = {378},
  pages = {686--707},
  year = {2019}
}

@article{Karniadakis2021SciML,
  author = {Karniadakis, George E. and Kevrekidis, Ioannis G. and Lu, Lu and Perdikaris, Paris and Wang, Sifan and Yang, Liu},
  title = {Physics-informed machine learning},
  journal = {Nature Reviews Physics},
  volume = {3},
  pages = {422--440},
  year = {2021}
}

@article{Cuomo2022ReviewPINN,
  author = {Cuomo, Salvatore and Di Cola, Vincenzo Schiano and Giampaolo, Fabio and Rozza, Gianluigi and Raissi, Maziar and Piccialli, Francesco},
  title = {Scientific Machine Learning through Physics-Informed Neural Networks: Where we are and what's next},
  journal = {Journal of Scientific Computing},
  volume = {92},
  pages = {88},
  year = {2022}
}

@article{Bradbury2018JAX,
  author = {James Bradbury and Roy Frostig and Peter Hawkins and others},
  title = {JAX: composable transformations of Python+NumPy programs},
  year = {2018},
  note = {http://github.com/google/jax}
}

@book{Griewank2008,
  author    = {Andreas Griewank and Andrea Walther},
  title     = {Evaluating Derivatives: Principles and Techniques of Algorithmic Differentiation},
  edition   = {2},
  publisher = {SIAM},
  year      = {2008}
}


\end{document}